\documentclass{webofc}
\usepackage[varg]{txfonts}   
\newcommand{\xmm}{{\em XMM-Newton}}
\newcommand{\xmms}{{\em XMM}}
\newcommand{\cxo}{{\em Chandra}}

\def \Rv {R_{500}}

\def\Mv {M_{\rm 500}}

\begin{document}
\title{SZ contribution to characterize the shape of galaxy cluster haloes}
%
%

\author{\firstname{Stefano} \lastname{Ettori}\inst{1,2}\fnsep\thanks{\email{stefano.ettori@inaf.it}}
 \and \firstname{Mauro} \lastname{Sereno}\inst{1} 
 \and \firstname{Sandra} \lastname{Burkutean}\inst{3}  
 \and \firstname{Jack}  \lastname{Sayers}\inst{4}
}

\institute{
INAF, Osservatorio di Astrofisica e Scienza dello Spazio, via P. Gobetti 93/3, 40129 Bologna, Italy 
\and
INFN, Sezione di Bologna, viale Berti Pichat 6/2, I-40127 Bologna, Italy
\and
INAF, Istituto di Radioastronomia -- Italian ARC, via P. Gobetti 101, 40129, Bologna, Italy
\and
California Institute of Technology, 200 E California Blvd., Pasadena, CA 91125
         }

\abstract{%
We present the on-going activity to characterize the geometrical properties of 
the gas and dark matter haloes using multi-wavelength observations of galaxy clusters.
The role of the SZ signal in describing the gas distribution is discussed for the pilot case of the CLASH object  MACS J1206.2-0847.
Preliminary images of the NIKA2 and ALMA exposures are presented.
}
\maketitle
\section{Introduction}
\label{intro}

As the most massive objects in dynamical equilibrium in the Universe, galaxy clusters represent fundamental signposts in the history of structure formation and evolution. The distribution of their dark and baryonic mass is the key ingredient to use galaxy clusters as astrophysical laboratories, cosmological probes, and tests for fundamental physics. The properties of the ongoing physical processes in galaxy clusters can be probed at different wavelengths, using galaxy distribution as observed in optical and infrared, X-ray surface brightness and spectral observations of the intra-cluster medium (ICM), the Sunyaev-Zeldovich effect (SZe) induced from the same ICM in the mm-band, and the gravitational lensing observations of the total mass distribution.

The analysis of these datasets requires some assumptions on the cluster geometry or hydrostatic equilibrium. 
An unbiased analysis has to take into account shape and orientation (e.g. \cite{limousin13}).
The cluster shape shows how matter aggregates from large--scale perturbations (\cite{west94, jing02}).
Assessing the equilibrium status is crucial to determine evolution and mechanisms of interaction of baryons and dark matter 
(\cite{lee03, kazantzidis04}).
Weak lensing (WL) analyses are independent of the equilibrium state and baryonic processes, but can measure only the projected mass. 
To infer the 3D mass, we have to deproject the lensing maps assuming a cluster shape, which is (commonly) ignored in the literature. 
The assumption of spherical symmetry introduces also biases in the measurement of the total mass and concentration 
(\cite{oguri05, sereno12}).

Sereno, Ettori et al. (2017 \cite{sereno17}) 
have started the CLUster Multi-Probes in Three Dimensions (CLUMP-3D) project to get the unbiased intrinsic properties of galaxy clusters
by exploiting rich datasets ranging from X-ray, to optical, to SZ wavelengths.
In a nutshell, lensing constrains the 2D mass and concentration which are deprojected thanks to the information on shape and orientation from X-ray (surface brightness and temperature) and SZ. 
The mass and concentration can be then determined together with the intrinsic shape and equilibrium status of the cluster as required by precision astronomy through a Bayesian inference method. 
In our modelling, we do not rely on the assumptions of spherical symmetry or hydrostatic equilibrium, which could bias results.
The joint exploitation of different data-sets improves the statistical accuracy and enables us to measure the 3D shape of the cluster's halo 
and any hydrostatic bias, evaluating the role of the non-thermal pressure support.

\begin{figure}[ht]
\includegraphics[width=0.45\textwidth]{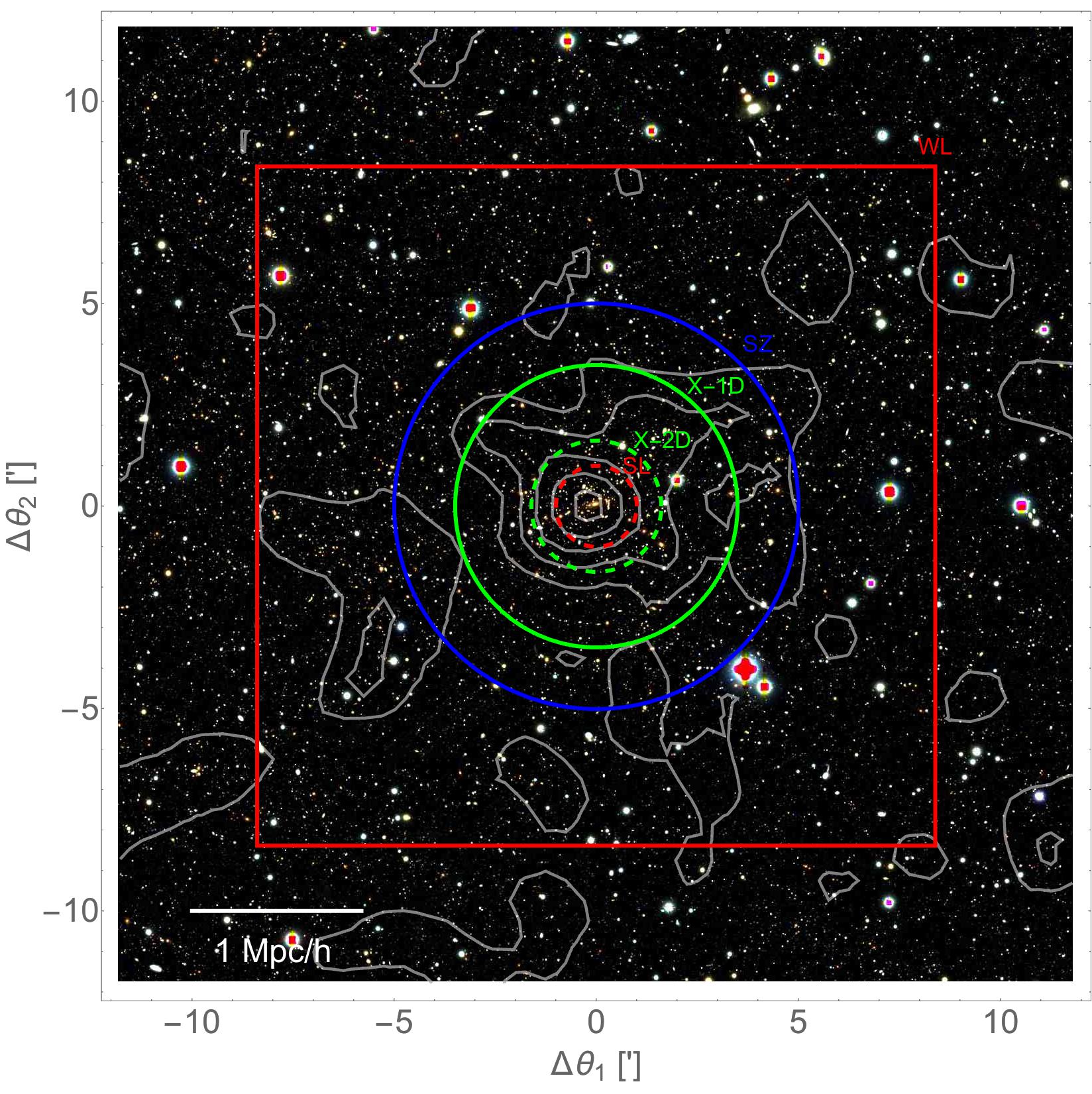}
\includegraphics[width=0.45\textwidth]{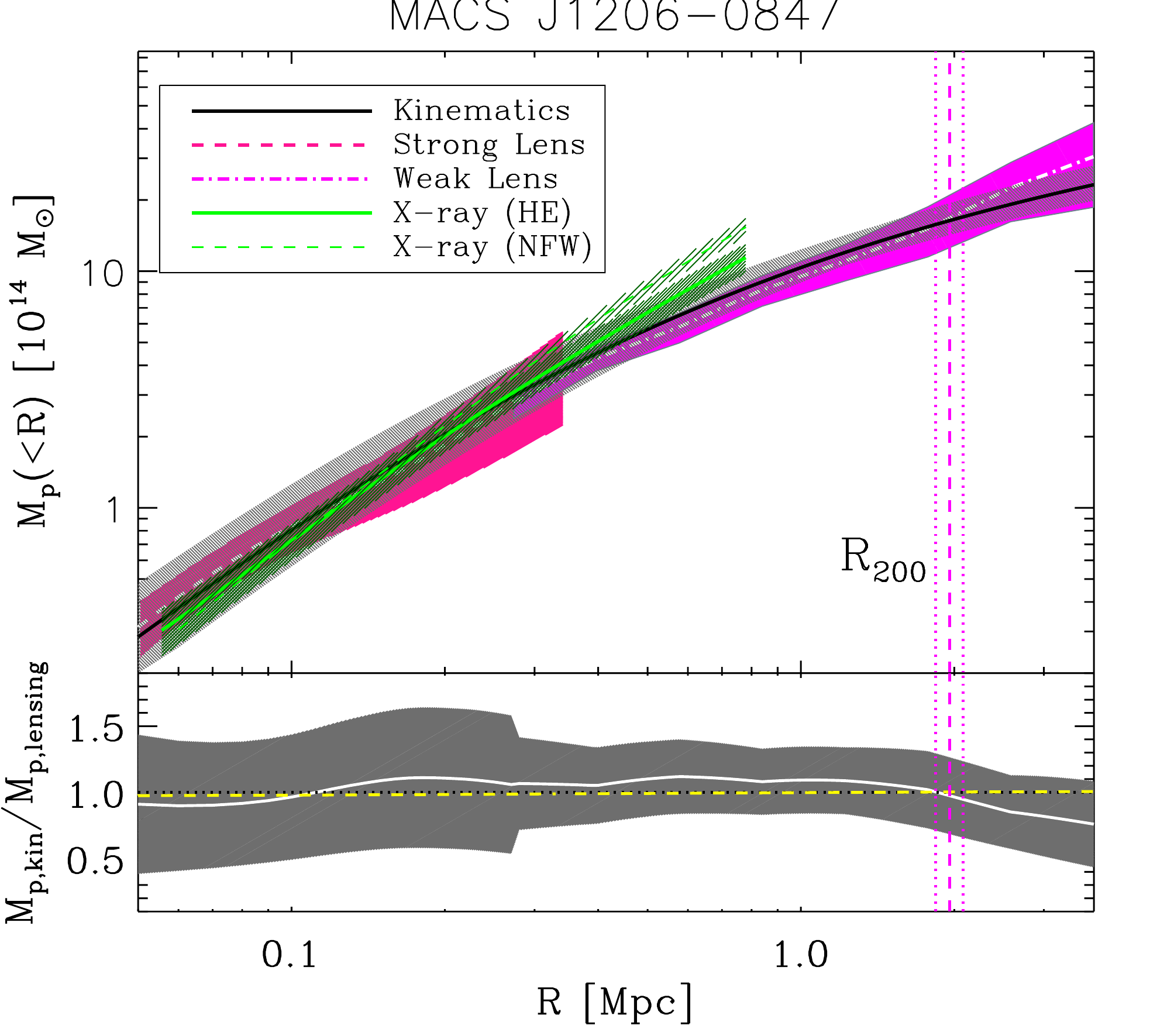}
 \caption{ {
\scriptsize 
{\bf (Left)}
Subaru $BVR$ composite colour images of the galaxy cluster MACS1206. 
The geometric forms centred in the optical centre enclose the regions exploited for inference by each probe (see details in \cite{sereno17}):
the dashed red circle at 1 arcmin and the red square of semi-size $8.39' \sim 2.8$ Mpc (1 arcmin corresponds to 0.34 Mpc $\approx$ 26\% of $R_{500}$)
for SL and WL; the dashed and full green circles at $\theta_{80\%}=1.61'$ and $3.49'$ for the 2D- and 1D-X-ray analysis, respectively; 
the blue circle at 5  arcmin for the SZe/Bolocam. 
The smoothed mass contours from the WL analysis of the Subaru observations are overlaid in white. 
The convergence levels go from $\kappa= 0.1$ to 0.5, with increments of 0.1. 
{\bf (Right)}
Compilation of the projected total mass profiles (lensing: Umetsu et al. 2012; kinematics: Biviano et al. 2013; 
more constraints in the inner 300 kpc from newly identified multiply lensed images are available in \cite{caminha17}.
They will be used to assess the robustness of the hydrostatic mass measurement.
The hydrostatic X-ray masses  (in green) are obtained by both assuming a NFW mass profile and applying directly the hydrostatic equilibrium equation
(e.g. \cite{ettori13}) to the deprojected gas density and poorly determined temperature profiles.
These two determinations encompass the range of systematic uncertainties from the current archival 22 ksec \cxo\ data. 
New proprietary (PI: Ettori) 180 ksec \cxo\ data will improve significantly the constraints on the temperature profile 
and will resolve with high S/N ratio the gas emission in the inner 30 arcsec.
The bottom panel shows the ratio of kinetic and lensing mass determinations; the horizontal dashed yellow line is the ratio of the corresponding
NFW fits, i.e. we find NFW$_{kin}\simeq$NFW$_{lens}$. 
}
} \label{fig:m1206}
\end{figure}

\begin{figure}[ht]
\includegraphics[width=0.9\textwidth]{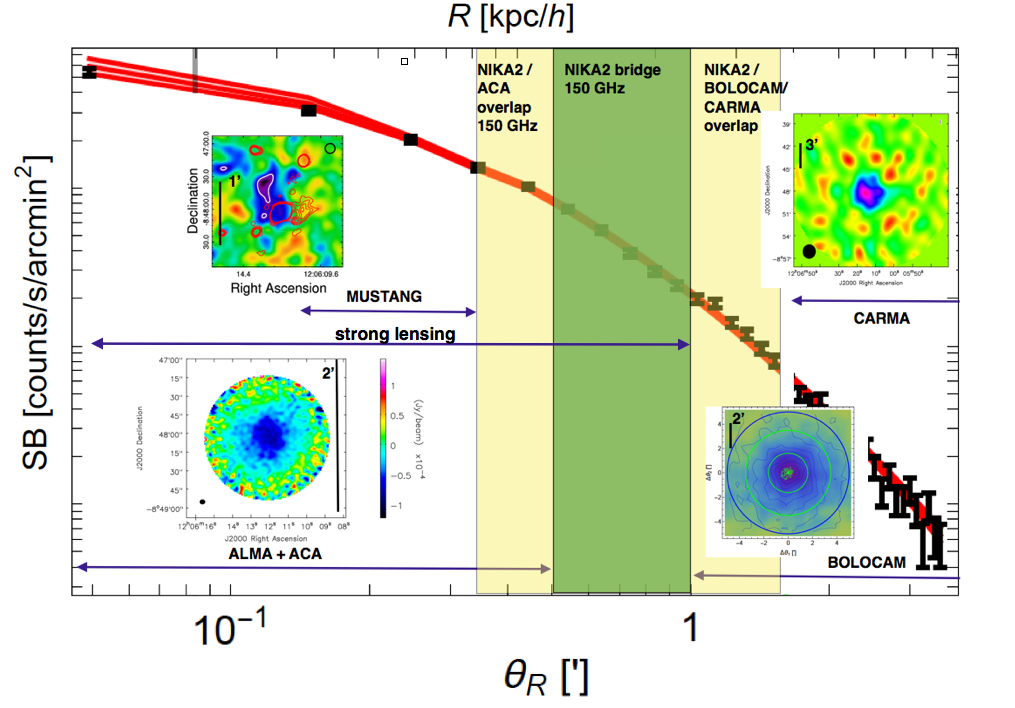}
 \caption{ {
\scriptsize Radial scales covered by all available SZ data. 
Arrows show the angular radial range from angular resolution to maximum recoverable radial scale for MUSTANG (\cite{young15}), 
our ALMA/ACA mock observation prediction, CARMA as well as Bolocam data. 
The yellow range shows the NIKA2 range of radial scales probed while the green shaded area illustrates the NIKA2 bridge between ACA and BOLOCAM/CARMA data. 
We plan to produce a combined multi-instrument SD+interferometric SZ map of these data, stressing the synergy between these different
instruments as well as perform  joint pressure profile model fitting taking into account the NIKA2 transfer function. 
Note that a similar SZ pressure profile fitting procedure using CARMA and APEX-SZ data was already carried out by one 
of the coI (Burkutean 2014  PhD thesis) on the galaxy cluster MACS0451.
}
} \label{fig:nika2}
\end{figure}

As a pilot study to address this fundamental science case, we aim at completing the wavelength coverage 
over different scales of the galaxy cluster MACS~J1206.2-0847 (see Fig.~\ref{fig:m1206}), 
for which the only missing piece in the puzzle is SZ at high angular resolution and over a wide radial scale (see Fig.~\ref{fig:nika2}).
Our joint analysis will offer a unique opportunity to obtain
an accurate and precise estimate of the baryonic and dark mass distributions, together with their 3D shapes, 
assessing at high accuracy any systematic bias affecting the measurements of the hydrostatic mass 
once it is compared with the dynamical and lensing masses from the inner core out to 2 Mpc.

\section{The case of MACS1206}
\label{sec-1}

MACS~J1206.2-0847 (\cite{sereno17}) is an X-ray luminous cluster at $z = 0.44$ 
that has HST broadband photometry as part of the CLASH (Cluster Lensing And Supernova survey with Hubble, 
\cite{postman12}) sample. 
Umetsu et al. (\cite{umetsu12}) performed an accurate cluster mass reconstruction 
from a combined weak-lensing distortion, magnification, and strong-lensing analysis of wide-field Subaru and HST imaging 
(see also \cite{merten15, umetsu16}).
The spectroscopic campaign for MACS1206 was completed in 2012 using VLT/VIMOS ({\it ESO LP}, PI: Rosati).  
Recently, the VLT/MUSE integral-field spectrograph has been used to augment the CLASH-VLT spectroscopic data set, 
particularly in the cluster core, resulting in a total sample of $\sim 700$ spectroscopic members (\cite{biviano13}).
This is one of the largest spectroscopic sample of cluster members available to date for a $z > 0.1$ cluster, and indeed at all redshifts.
With the aforementioned VLT/MUSE observations, a large number of new multiply lensed images have been identified 
and used to build a high-precision lensing model (\cite{caminha17}).
{\sl The combination of all these independent methods has led to the most
accurate ever determination of the total gravitating mass for an individual cluster over the radial
range 5-5000 kpc (corresponding to about 0-2.5 virial radii; see Fig.~\ref{fig:m1206})}. 

This deep optical coverage is completed by X-ray data (180 ksec awarded in last \cxo\ AO, PI: Ettori, 
to resolve with high S/N ratio the gas emission in the inner 30 arcsec)
and ancillary measurements of the SZe collected with Bolocam (\cite{czakon15}), CARMA and 
ACA + ALMA 12m ({\it proposal 2017.1.01430.S awarded with 14.3 hours in the ALMA Cycle 5}; PI: Ettori). 
This large dataset available for such a remarkably regular, face-on, massive cluster makes MACS1206 an ideal target for a detailed multi-wavelength analysis.
The CLUMP-3D analysis has allowed to reconstruct the triaxial shape of the cluster, showing that
the main axis is near the plane of the sky, and that the thermal pressure can balance the cluster in hydrostatic equilibrium.
Improving the quality of the X-ray and SZ dataset allows to derive with similar accuracy all the baryonic mass components and the
dark matter distribution thus {\it probing fundamental predictions of the CDM model:  the shape of a cluster's dark-matter potential 
and the baryon-to-dark matter ratio}. Our high-resolution SZ (ALMA 12m, ACA 7m and MUSTANG) data will map the SZ 
signal at angular resolutions from 4 to 9 arcsec up to radial scales of 0.5 arcmin. 
Our low resolution Bolocam and CARMA data cover angular radial scales from 1 to 5 arcmin at 1 arcmin angular resolution. 
We are therefore missing high resolution SZ data at radial scales from 30 to 100 arcsec. 
NIKA2  at 150 GHz is the ideal instrument to provide us with this missing piece in the puzzle (see Fig.~\ref{fig:nika2}). 
Its 17.7 arcsec angular resolution will give three independent beams within radial scales of 1' and thus bridge the gap between the ACA and Bolocam data, 
while covering the GNFW model scale radius (\cite{nagai07}) at $51\pm 2$ arcsec determined from strong lensing  (\cite{caminha17}).
The joint analysis of the SZ effect from 3.3 arcseconds (ALMA) to 5 arcminutes (IRAM/NIKA2, Bolocam) will enable us
to make it comparable to our X-ray data, allowing e.g.
(i) to constrain the temperature profile by combining the thermal pressure from SZ and the gas density from X-ray observations 
(as also highlighted in \cite{ruppin17} and \cite{adam17},
and (ii) to look for radial variations in the gas shape. 
NIKA2 data is vital for a complete radial coverage. 

\begin{figure}[ht]
\hbox{
\includegraphics[width=0.6\textwidth]{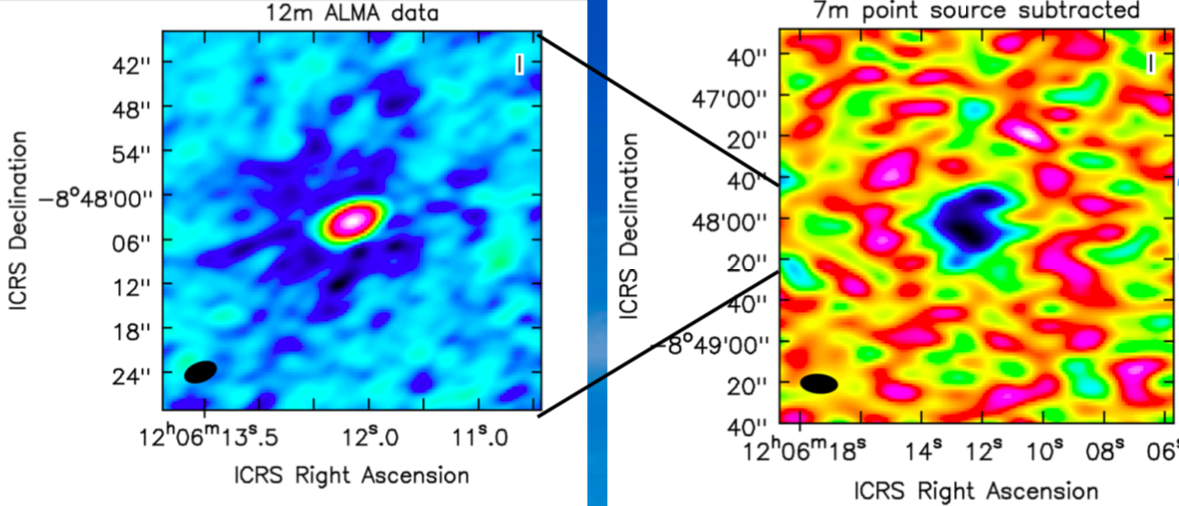}
\hspace*{0.3cm} 
\includegraphics[width=0.28\textwidth]{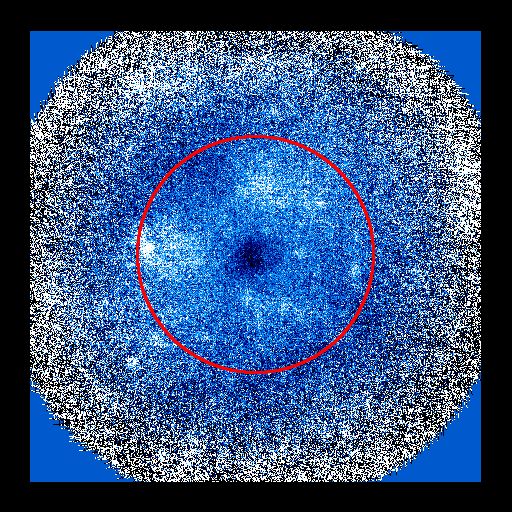}
}
 \caption{Products of the preliminary data analysis.
ALMA 12m and 7m images (briggs weighting robust: 2.0) with the point source removed using a single point source clean component model. 
The clean component is based on the images of the 12m data that upweight the long baselines.
(Right) NIKA2 2mm image as provided by B. Ladjelate and IRAM+NIKA2 teams. The red circle has radius equal to $R_{500} \approx 3.9$ arcmin.
} \label{fig:prel}
\end{figure}

In Figure~\ref{fig:prel}, we present very preliminary images from our proprietary ALMA and NIKA2 exposures.

\section{3D analysis in the context of the Heritage \xmm\ Cluster Project}

In \xmm\ AO17 (2017) call, we have been awarded with 3 Msec to complete homogeneous and systematic X-ray exposures of 118 Planck-SZ selected galaxy clusters comprising an unbiased census of (i) the population of clusters at the most recent time ($z < 0.2$), (ii) the most massive objects to have formed thus far in the history of the Universe at $z<0.6$. Details and updated information on this  \xmms\ Multi-Year Heritage Project, titled ``{\it Witnessing the culmination of structure formation in the Universe}''  (PI: M. Arnaud -CEA Saclay- and S. Ettori -INAF OAS Bologna) are available at {\it xmm-heritage.oas.inaf.it}.

The \xmms\ exposure times are set to reach a $S/N=150$ that will allow us to map the temperature profile  in 8+ annuli at least up to $\Rv$, with a precision of  
$\pm15\%$  in the $[0.8$--$1.2]\Rv$ annulus, to reach an uncertainty of $\pm 2 \%$ on $\Mv^{YX}$, and to derive the hydrostatic mass measurements at $\Rv$ to the $\sim15$--$20\%$ precision level. 

\begin{figure}[ht]
\includegraphics[width=0.9\textwidth]{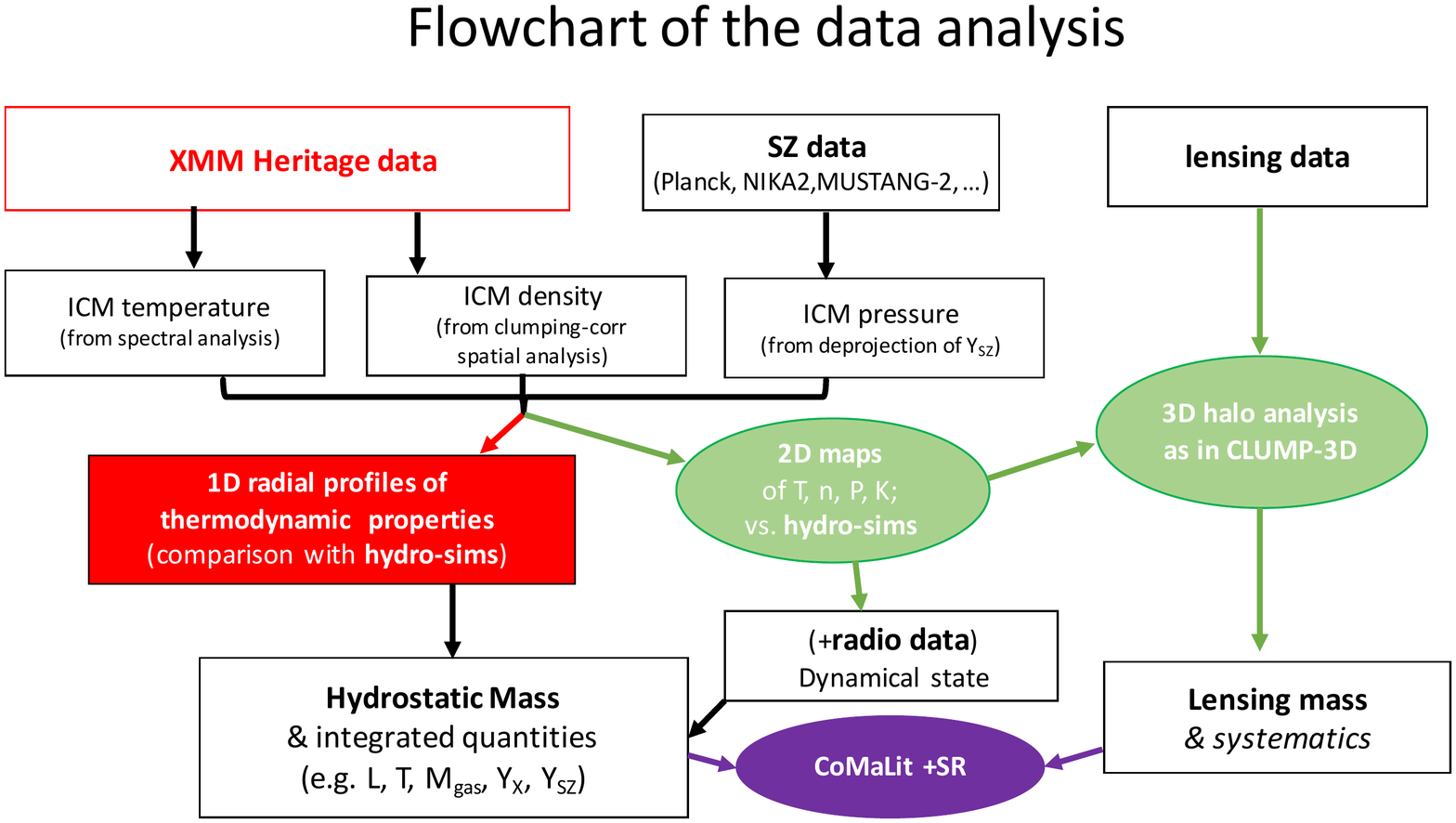}
 \caption{ {
Flowchart of the (possible) data analysis for the Heritage \xmm\ cluster project.
"CoMaLit +SR" refers to the analysis of the scaling relations (SR) in the {\it CoMaLit} framework (see for details e.g. \cite{sereno17b}), where we adopt a robust Bayesian technique to characterize both the bias between hydrostatic and lensing mass estimates and the properties of the observables-mass relations, inferring also their mass and redshift dependence, the time-evolving intrinsic scatter, and correcting by any modeled observational bias.
}
} \label{fig:analysis}
\end{figure}

As represented in a flowchart in Fig.~\ref{fig:analysis}, 
we will make spectral and spatial multi-dimensional analyses to assess the systematics affecting our 1D radial profiles. 
Thermodynamic 2D maps have been extensively used in the study of galaxy clusters thanks to their great potential to characterize the dynamical state of a system. 
We will use these maps to measure the fluctuations and scatter present in the azimuthally-averaged profiles, 
isolating regions that induce anomalies in the 1D profiles, like infalling clumps and ongoing mergers (e.g. \cite{ghirardini18}).
Unrelaxed clusters are expected to show stronger (and possibly correlated) fluctuations in the gas temperature and density distribution (e.g. \cite{rasia14}). Therefore, any diagnostic of substructures is expected to be linked to the scatter of the scaling relations, through biases in the hydrostatic mass and in the integrated X-ray quantities. Multi-probes three-dimensional Bayesian analysis will be applied to combined X-ray, SZ and lensing data to fit three-dimensional triaxial ellipses to the gas and total mass distribution, as we probed for 16 X-ray regular CLASH clusters (CLUMP-3D project; \cite{sereno18}). 
In general, we obtained that the shapes are in good agreement with the predictions from the standard $\Lambda$CDM cosmological model. 
However, compared to simulations, the data show a slight preference for more extreme minor-to-major axial ratios. We need a combination of more sensitive observational data and a larger cluster sample, as available in the Heritage project, to probe, also as function of halo mass and dynamical state, the 3D structure of the gas and dark matter distribution, assessing their consistency with $\Lambda$CDM predictions and their equilibrium once geometrical biases (like projection) are corrected for.

%
%

\end{document}